# Circuit Modeling of Tunneling Real-Space Transfer Transistors: Toward Terahertz Frequency Operation

Wen Huang, *Student Member, IEEE*, Xin Yu, *Student Member, IEEE*, Shi-Lin Zhang, Lu-Hong Mao, and Jean-Pierre Leburton, *Fellow, IEEE*

*Abstract*—High frequency operation of tunneling real-space transfer transistor (TRSTT) in the negative differential resistance (NDR) regime is assessed by calculating the device common source unity current gain frequency ($f_T$) range with a small signal equivalent circuit model including tunneling. Our circuit model is based on an $In_{0.2}Ga_{0.8}As$ and δ-doped GaAs dual channel structure with various gate lengths. The calculated TRSTT $f_T$ agrees very well with experimental data, limiting factor being the resistance of the δ-doped GaAs layer. By optimizing the gate dimensions and channel materials, we find $f_T$ in the NDR region approaches terahertz range, which anticipates potential use of TRSTT as terahertz sources.

*Index Terms*—Tunneling real-space transfer transistor (TRSTT), negative differential resistance (NDR), small signal equivalent circuit model, unity current gain frequency

## I. Introduction

TUNNELING real space transfer (TRST) is a quantum mechanical effect that arises between two channels of different mobilities in field effect transistors to achieve NDR controlled by the gate bias [1]. The effect was first demonstrated in pseudo-morphic AlGaAs/InGaAs MODFET [2], and investigated in several TRSTT structures in the last decades [3]–[5]. Recently, very clean TRST-induced NDRs with modulated peak to valley (P/V) ratios up to 4 at room temperature were demonstrated in pseudo-morphic GaAs /InGaAs MODFET structures (fig.1 inset) [6]. The devices however were relatively long and wide with large capacitance, so that $f_T$ was below 10 GHz. Therefore considerable room for improvement is expected with frequency operation far above 100 GHz, and hopefully up to the THz range [7].

The TRSTT operation frequency in the NDR region depends on two main factors: the tunneling time $\tau_t$ between the two channels, and the carrier transit time $\tau_{SD}$ from the source to drain in the dual channel structure. Usually, the tunneling time is estimated to be less than 0.5 ps [6], so the TRSTT performances are essentially determined by $\tau_{SD}$.

In this letter, based on devices similar to those in ref [6] we implement a small-signal equivalent circuit model accounting for tunneling between the TRSTT two channels, and for which $f_T$ in tunneling mode is calculated analytically in the common source circuit configuration. Our model predicts that $f_T$ in NDR region can reach THz range by optimizing the channel material, and shortening the gate length.

## II. TRSTT Structure and Measurement

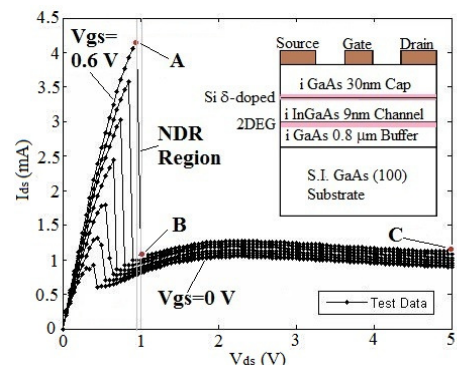

Fig. 1. Experimental drain current vs. drain voltage under different gate voltages in TRSTT starting from $V_{gs}$=0 V to $V_{gs}$=0.6 V with a step of 0.1 V.

The device structure is schematically shown in fig.1 inset. The heterostructures consist of a 0.8-μm intrinsic GaAs buffer layer, a 90-Å undoped $In_{0.2}Ga_{0.8}As$ channel layer, a 90-Å undoped GaAs spacer layer, followed by a silicon δ-doped layer with a $4\times10^{12}$ cm$^{-2}$ sheet density, and a 300-Å undoped GaAs cap layer. The gate dimension is $2\times60$ μm$^2$, and the source-gate and gate-drain separations are 2 μm, each.

Fig. 1 displays the device output characteristics with NDR

Manuscript received December 7, 2011. This work is supported in part by the Beckman Institute for Advanced Science and Technology at the University of Illinois at Urbana-Champaign, the School of Electronic Engineering at the University of Electronic Science and Technology of China, and Tianjin University.

Wen Huang is with Beckman Institute for Advanced Science and Technology at the University of Illinois at Urbana-Champaign, Urbana, IL, 61801, USA and the School of Electronic Engineering at the University of Electronic Science and Technology of China, Chengdu, Sichuan, 611731, China (e-mail:whuang82@illinois.edu).

Xin Yu is with the Department of Electrical and Computer Engineering at University of Illinois at Urbana-Champaign, Urbana, IL 61801 USA. (e-mail: xinyufisher@gmail.com).

Lu-Hong Mao and Shi-Lin Zhang are with the Department of Electronics and Information Engineering, Tianjin University, Tianjin 300072, China.

J. P. Leburton is with the Department of Physics, Department of Electrical and Computer Engineering, and the Beckman Institute for Advanced Science and Technology at the University of Illinois at Urbana-Champaign, Urbana, IL, 61801, USA (phone: 217-333-6813; e-mail: jleburto@illinois.edu).

under different gate biases. Noticeable gate leakage current is observed with the NDR onset, which is attributed to TRST [6]. The channel electron mobility is measured to be 5604 cm²/Vs by Hall test, and the sheet density of carrier in $In_{0.2}Ga_{0.8}As$ channel is $9.02\times10^{11}$ cm$^{-2}$ at room temperature. In the absence of NDR i.e. TRST, $f_T$ at bias point C ($V_{gs}$=0.6V, $V_d$=5V) is measured to be 8.9 GHz[1].

## III. SMALL SIGNAL EQUIVALENT CIRCUIT WITH TUNNELING

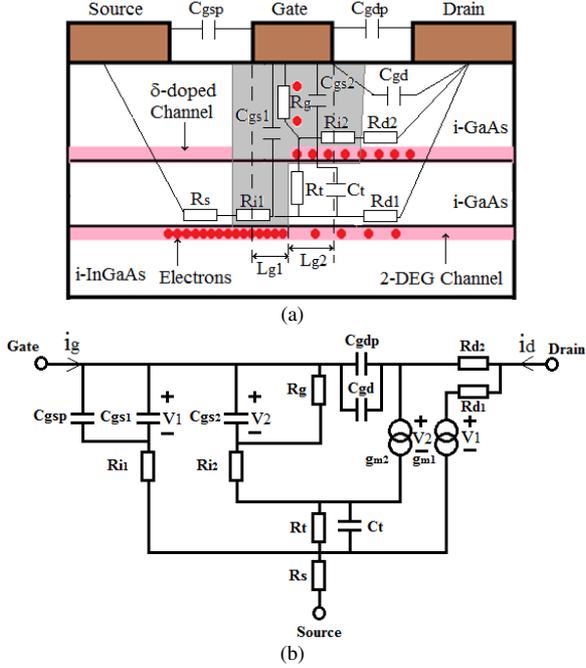

Fig. 2. TRSTT small-signal equivalent circuit in the common source configuration . (a). Functions of each circuit element with corresponding position. (b). Schematic of the equivalent circuit.

In Fig.2 (a) we show the small-signal equivalent circuit of the TRSTT including tunneling effect. Under specific combination of gate and source-drain biases, most of electrons tunnel from the high mobility 2-dimensional electron gas (2-DEG) channel in the InGaAs material to the low mobility silicon δ-doped channel [1]–[2], where few electrons leak to the gate leakage. The TRSTT behavior in the tunneling mode can be described as two enhancement HEMTs operating in series with the same gate bias. In the first one the 2-DEG is in the InGaAs channel and in the other one it is in the silicon δ-doped channel. By following the paths the electrons flow, one can build the equivalent circuit shown in fig.2 (b). $C_{gsp}$ and $C_{gdp}$ are source-gate and gate-drain geometrical capacitances whose values are determined by the dimensions of electrode layout. $C_{gs1}$ and $C_{gs2}$ are the gate-source intrinsic capacitances that control the source-drain current, while $g_{m1}$ and $g_{m2}$ are the corresponding transconductances. $C_{gd}$ is the drain-source intrinsic capacitance which describes the electron inflow into the depletion layer when the electrode voltage changes. $R_{i1}$ and $R_{i2}$ are the dual channel resistances, while $R_{d1}$ and $R_{d2}$ are the drain parasitic resistances for the 2-DEG InGaAs channel and the silicon δ-doped channel, respectively. $R_s$ is the total source parasitic resistance for the two channels. $R_t$ and $C_t$ are the tunneling resistance and capacitance, with their product yielding the tunneling time. In our model, we assume tunneling between the two channels occurs at a certain point right under the gate in the 2-DEG channel. We define $L_{g1}$ and $L_{g2}$ as the effective lengths of the 2-DEG and δ-doped channels through which TRST electrons transit (see fig.2.a). Then $L_{g1}/L_g$ determines the tunneling position under the gate. As the tunneling distance is short, only a fraction of gate voltage drops across the tunneling barrier, although generating a significant tunneling current, and correspondingly a negligible tunneling resistance $R_t$ compared to the undoped GaAs cap layer.

## IV. UNITY CURRENT GAIN FREQUENCY IN THE NDR REGION

$f_T$ is defined when $|i_d/i_g| = 1$. For the sake of simplicity, $C_{gd}$ and $C_{gdp}$ are ignored here, because of their small values. From the small-signal form of Kirchhoff's circuit laws one gets for the drain current, the gate current, and the voltage drops across $C_{gs1}$ and $C_{gs2}$ ,respectively.

$$i_d = g_{m1}v_1 + g_{m2}v_2 \quad (1)$$

$$i_g = j\omega(C_{gs1} + C_{gsp})v_1 + j\omega C_{gs2}v_2 + \frac{v_2}{R_g} \quad (2)$$

$$v_1 = \frac{v_g - (i_g + i_d)R_s}{1 + j\omega(C_{gs1} + C_{gsp})R_{i1}} \quad (3)$$

$$v_2 = \frac{v_g - (i_g + i_d)R_s}{1 + \frac{R_{i2} + \frac{R_t}{1+j\omega\tau_t}}{R_g} + \frac{g_{m2}R_t}{1+j\omega\tau_t} + j\omega C_{gs2}\left(R_{i2} + \frac{R_t}{1+j\omega\tau_t}\right)} \quad (4)$$

with $\omega = 2\pi f_T$, $\tau_t = R_t C_t$. Owing to the strong I vs. V non-linearity, the $f_T$ determination in the NDR region would be tedious under small signal operation. Rather, one can estimate its value by calculating it on both sides of the NDR region i.e. at points A and B shown on fig. 1(a). By ignoring the gate leakage, one obtains the following equation for drain currents in the two channels right before and after tunneling,

$$n_{s1}qW_g v_{e1} + n_{s2}qW_g v_{e2} = \frac{I_{dB}}{I_{dA}}(n_{s1} + n_{s2})qW_g v_{e1} \quad (5)$$

Here q is the electron charge, $I_{dB}$ and $I_{dA}$ are the drain currents at bias points A and B, $v_{e1}$, $v_{e2}$ are the saturation velocity of electrons in both channels, $W_g$ is the gate width, and $n_{s1}$ and $n_{s2}$ are the carrier density in the 2-DEG channel and δ-doped channel under the gate after tunneling. Based on the definition $g_{m1}$ and $g_{m2}$, one gets

$$g_{m1} = \frac{\gamma_P - \gamma_V}{\gamma_P(1-\gamma_V)}g_m, g_{m2} = \frac{\gamma_V(1-\gamma_P)}{\gamma_P(1-\gamma_V)}g_m \quad (6)$$

---
[1] In ref. [6], the mentioned $f_T$=9 GHz is taken from the 6 μm channel length sample used in our modeling and its accurate value is 8.9 GHz.

where we call $\gamma_P = I_{dB}/I_{dA}$ and $\gamma_V = v_{e2}/v_{e1}$ the tunneling peak ratio and velocity difference ratio.

Usually, the gate-source capacitance $C_{gs}$ is defined as the net increase of positive charge in the depletion area by an incremental increase in gate-source voltage [8]. However for δ-doped TRSTT, the only positive charge is in the δ-doped layer. Instead, we obtain $C_{gs}$ indirectly by calculating the variation of the carrier densities $Q_1$ and $Q_2$ in the two channels under the gate. In the tunneling mode, the total gate-source capacitance can be written as

$$C_{gs} = \frac{\partial Q_1}{\partial V_{gs}} + \frac{\partial Q_2}{\partial V_{gs}} = qL_{g1}W_g\frac{\partial n_{s1}}{\partial V_{gs}} + qL_{g2}W_g\frac{\partial n_{s2}}{\partial V_{gs}} \quad (7)$$

The drain currents in the two channels read

$$I_{d1} = qW_g v_{e1} n_{s1} = g_{d1}V_{ds}, I_{d2} = qW_g v_{e2} n_{s2} = g_{d2}V_{ds} \quad (8)$$

where $g_{d1}$, $g_{d2}$ are the drain conductance of the 2-DEG and δ-doped channels, respectively. Combining (7), (8), one can get

$$C_{gs} = \frac{L_{g1}}{v_{e1}}g_{m1} + \frac{L_{g2}}{v_{e2}}g_{m2} \quad (9)$$

One determines $f_T$ in the tunneling mode by solving (1)–(4), (6) and (9). In a first order approximation one can neglect $R_{i1}$, $R_{i2}$, $R_s$, $R_d$, $C_{gsp}$, and $\tau_t$ compared to other circuit elements to obtain a closed form for $f_T$ at point B as shown in (10)

$$f_{T\_B} = \frac{v_{e1}\gamma_P(1-\gamma_V)}{2\pi[L_{g1}(\gamma_P - \gamma_V) + L_{g2}(1-\gamma_p)]} \quad (10)$$

When no tunneling occurs, $\gamma_P$=1 and $\gamma_V$=0. The $f_T$ at point A can be derived from that of $f_{T\_B}$ as shown in (11)

$$f_{T\_A} = \frac{v_{e1}}{2\pi L_g} \quad (11)$$

## V. RESULTS AND DISCUSSION

Based on the measured channel mobility, one can estimate the saturation velocity $v_{e1}$ of the sample to be $1\times10^7$ cm/s. Inserting this value into (11), we get $f_T$=7.96 GHz, which agrees well with the experimental value 8.9 GHz obtained for the device in saturation, in the absence of TRST[1].

In Fig. 3, we show the calculated $f_T$ as a function of gate length around the NDR region for different tunneling positions along the 2-DEG channel. The best $f_T$ can be achieved by shrinking the gate length to 40 nm with a tunneling position close to the drain. If we optimize the saturation velocity to $2\times10^7$ cm/s by using higher indium composition InGaAs material and maintain the values of other parameters in (10) and (11), $f_T$ in the NDR region can reach a range between 620 GHz and 800 GHz for $L_{g1}/L_g$=0.7, which indicates that high frequency response is obtained for tunneling closer to the drain than the source, in order to reduce the δ-doped channel resistance. This range of $f_T$ is shown in fig. 3 between the black solid line with solid circles and the red solid line.

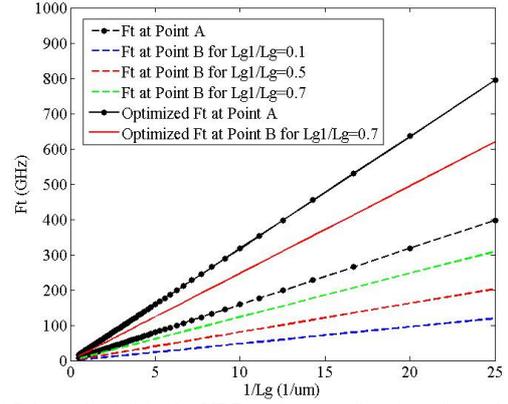

Fig. 3. (*Color online*) $f_T$ in the NDR region as a function of gate lengths for different tunneling positions and the optimized channel saturation velocities.

## VI. CONCLUSION

Our small signal equivalent circuit model for TRSTT shows that THz frequency operation in the NDR region of the devices is possible by shrinking the gate length, but also improving the mobility or saturation velocity in the 2DEG channel. As expected the critical factor is the δ-doped channel resistance that should remain significant compared to the 2DEG channel resistance if large NDR peak-to-valley ratios are required. In this context, it may be worth exploring other material systems offering optimum mobility/saturation velocity difference with large and abrupt NDRs. Finally, let us mention that a key assumption in our analysis was the TRST occurrence along the channel. For this purpose a physical model of quantum tunneling between channels under bias conditions is desirable.

## VII. ACKNOWLEDGEMENT

We are indebted to Dr. Elyse Rosenbaum for helpful discussions.